\documentclass[aps,prl,showpacs,amssymb,floatfix,twocolumn]{revtex4}
\usepackage{graphicx}
\usepackage{amssymb}

\begin{document}

\title{Mixing and segregation rates in sheared granular materials}
\author{Laura A. Golick and Karen E. Daniels}
\email{karen_daniels@ncsu.edu}
\affiliation{Department of Physics, North Carolina State University, Raleigh, NC, 27695}
\date{\today}

\begin{abstract}
The size-segregation of granular materials, a process colloquially known as the Brazil Nut Effect, has generally been thought to proceed faster the greater the size difference of the particles. We experimentally investigate sheared bidisperse granular materials as a function of the size ratio of the two species, and find that the mixing rate at low confining pressure behaves as expected from percolation-based arguments. However, we also observe an anomalous effect for the re-segregation rates, wherein particles of both dissimilar and similar sizes segregate more slowly than intermediate particle size ratios. Combined with the fact that increasing the confining pressure significantly suppresses both mixing and segregation rates of particles of dissimilar size, we propose that the anomalous behavior may be attributed to a species-dependent distribution of forces within the system.
\end{abstract}

\maketitle

Accurate knowledge of the rate at which granular materials segregate by size under shear \citep{Williams-1976-SPM,Ottino-2000-MSG} is significant for such applications as avalanche hazard prediction and the design of industrial particle separation chutes. One of the most common segregation phenomena, the Brazil Nut Effect, is broadly observed and has been associated with a  variety of proposed mechanisms \citep{Schroter-2006-MSS}. For shear flows, kinetic sieving theory in various forms \citep{Bridgwater-1976-FPM, Savage-1988-PSS, Gray-2005-TPS} has been the most promising. These theories rely on statistical arguments which quantify the creation of voids through shear: smaller particles preferentially fall into these voids in a percolation-like fashion. Therefore, it is expected that the larger the difference in particle sizes, the quicker this process will happen. While percolation rates have previously been measured in a quasi-two-dimensional experiment \citep{Scott-1975-IPF}, there is to date no fundamental understanding of the size-dependence and pressure-dependence for true three-dimensional flows, nor is it known in which regimes kinetic sieving is the dominant effect. 

We investigate the mixing and subsequent re-segregation of a granular material initially configured so that a layer of small particles is placed above an equal volume of large particles within an annular shear cell. Under shear from the bottom plate, the small particles migrate to the bottom and the large particles correspondingly migrate to the top, as is expected for particles of otherwise identical material \citep{Ottino-2000-MSG}. We measure the mixing and segregation rates as a function of particle size ratio and confining pressure and find that the mixing rate is consistent with kinetic-sieving models for approximately-hydrostatic confining pressure. However, the segregation rates are observed to be non-monotonic in particle size ratio, in contrast with kinetic-sieving theory, and strongly depend on the confining pressure. Below, we quantify these rates and interpret them in light of the heterogeneous force-transmission properties of granular materials.

\begin{figure}
\centerline{\includegraphics[width=\linewidth]{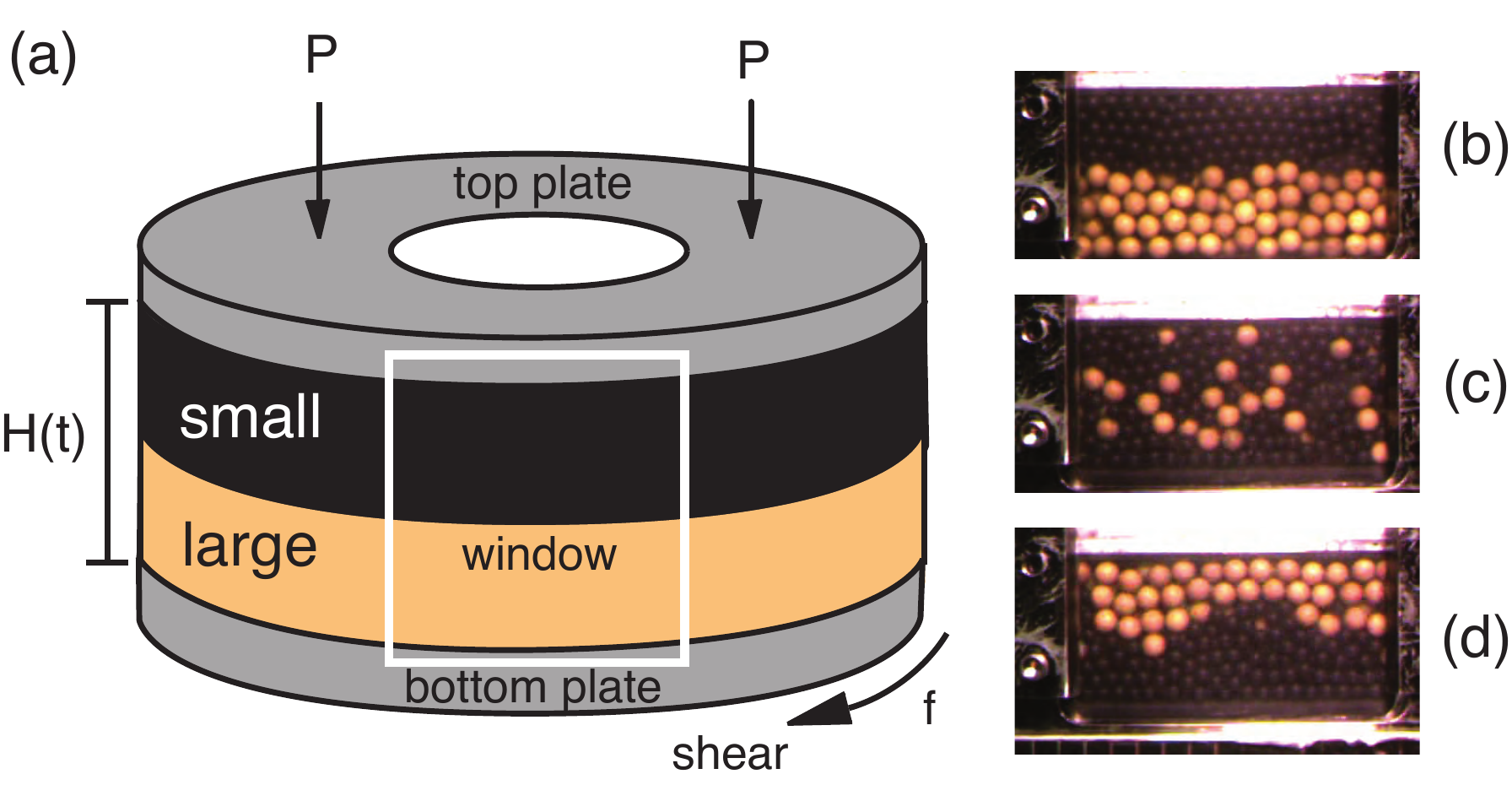}}
\caption{(a) Schematic of experimental apparatus (not to scale) showing initial configuration of particles within the annulus. Sample images taken at window for $d_S = 4$ mm (dark particles) and $d_L = 6$ mm (light particles): (b) initial configuration, (c) mixed state and (d) final re-segregated state. \label{f:exp}}
\end{figure}

Our experimental cell is an annulus which confines the particles between a top plate free to move vertically and a rotating bottom plate, as shown in Fig.~\ref{f:exp}(a). The radius of the inner wall of the annulus is $R = 292$ mm, and the width of the cell is 25.4 mm.
Both the top and bottom plates are lined with rubber to increase their friction coefficient. The bottom plate has a rotation period of $20.4$ seconds (frequency $f = 49.0$ mHz), creating a shear band which extends a few particle diameters \citep{May-2009-SDP} into the cell. 
We adjust the confining pressure of the cell via two techniques: weighting the top plate to increase the pressure, or partially suspending the top plate from springs to reduce the pressure on the granular aggregate. 

Each experimental run begins in an initial state consisting of a layer of small particles (mass 2 kg) over a layer containing an equal mass of large particles. This initial configuration is shown in Fig.~\ref{f:exp}(b). The small particles are a single size for each run, with diameter $d_S$ ranging from $1.5$ to $5.0$ mm; large particles are fixed at diameter $d_L = 6$ mm for all runs. Once shear begins, the small particles filter downwards through the large particles, resulting in a mixed state such as the one shown in (c). Eventually, nearly all of the large particles have reached the top of the cell (d). This experimental protocol allows us to examine both the mixing of small and large particles, and the subsequent re-segregation of the mixture.

\begin{figure}
\centerline{\includegraphics[width=\linewidth]{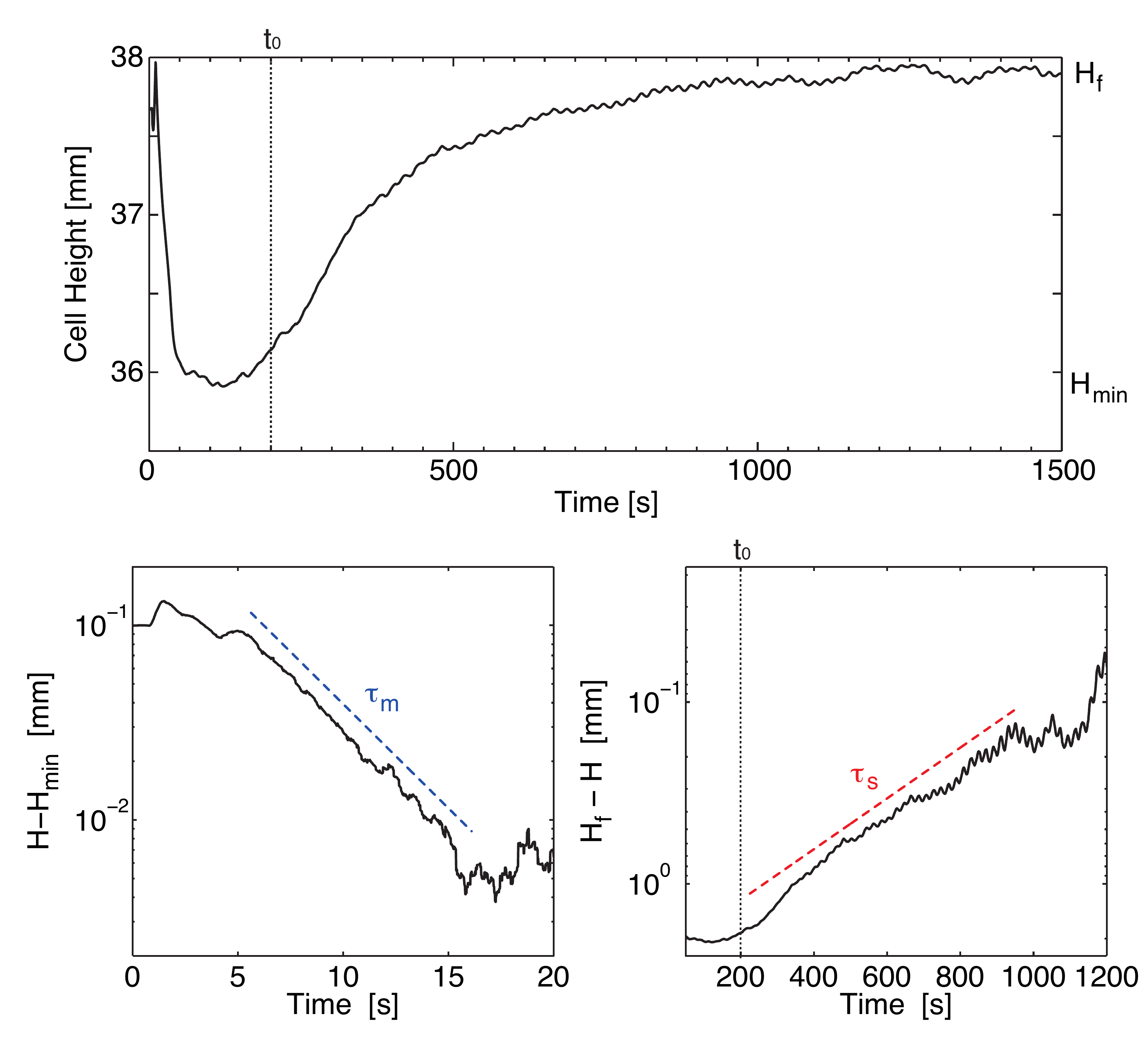}}
\caption{Sample cell height data, for $r=2/6$ and $\tilde P = 0.36$. An averaging window of 0.5 sec was used to smooth the raw signal. (a) Cell height $H(t)$, with values $H_\mathrm{min}$, $H_\mathrm{f}$, and $t_0$ marked. (b) Magnified portion of $H(t) - H_\mathrm{min}$ showing fit to determine mixing timescale $\tau_m$. (c) Magnified portion of $H_\mathrm{f} - H(t)$ showing fit to determine segregation timescale $\tau_s$. \label{f:data}}
\end{figure}

The view at the outside wall is not necessarily representative of the bulk behavior, particularly for mixtures of very different sized particles. Therefore, to measure the average behavior of the whole system, we monitor the height $H(t)$ of the top plate, as shown in Fig.~\ref{f:data}(a). When shear begins, the system initially expands due to Reynolds dilatancy. As small particles fill the gaps between the large particles during mixing, the overall cell height quickly decreases. When the particles are well-mixed, the aggregate takes up the least total space and falls to a height $H_\mathrm{min}$. As re-segregation occurs, the system re-dilates to a final height $H_\mathrm{f}$. We measure the timescale $\tau_m$ for this mixing process by fitting the function $H(t) - H_\mathrm{min} \propto e^{-t/\tau_m}$ to the decrease in the height of the cell. As the particles begin to re-segregate, the large particles rise through the mix and ultimately end up in a layer above the small particles. During this process, we define a segregation timescale $\tau_s$ by fitting a function of the form $H(t) - H_\mathrm{f} \propto e^{-(t - t_0)/\tau_s}$ where $t_0$ is chosen to be after the minimum $H_\mathrm{min}$. Representative fits for $\tau_m$ and $\tau_s$ are shown in Fig.~\ref{f:data}(b,c).

As a consequence of fixing the mass (volume) of the particles, the total number of particles varies with particle size ratio $r = d_S / d_L$. Therefore, we scale the cell height by an appropriate mean particle diameter $\delta$ such that $H/\delta$ represents the height of the cell measured in particle diameters. For each $(d_S, d_L)$ pair, we define $\delta$ via the relationship
\begin{equation}
\frac{2}{\delta} = \frac{1}{d_S} + \frac{1}{d_L}. 
\label{delta}
\end{equation}
Using $\delta$, we can properly compare non-dimensionalized mixing ($\Omega_m$) and segregation ($\Omega_s$) rates among runs with different $r$: 
\begin{equation}
\Omega_{m,s} = \frac{H}{\delta f \tau_{m,s}}
\label{rate}
\end{equation}
where $f$ is the rotation frequency of the bottom plate. 

By varying the confining pressure, we explore a regime both above and below an approximately ``hydrostatic'' pressure due to the weight of the particles. We scale the effective weight of the top plate by the weight of the particles, and report dimensionless $\tilde P$ defined as: 
\begin{equation}
\tilde P  = \frac{m_\mathrm{p} g + M g - k \, \Delta x}{m_\mathrm{g} g}
\label{P}
\end{equation}
where $m_\mathrm{p} = 15.42$ kg is the mass of the top plate, $m_\mathrm{g} = 4$ kg is the total mass of the particles, $M$ is the added compressive mass (if present), and $k \, \Delta x$ is the average upward force from the supporting springs (if present). We explore values of $\tilde P$ from 0.25 to 1.48; typical variation within a single run is $\pm 8 \times 10^{-3}$ due to the contraction/extension of the supporting springs. We add mass $M = 0$ to 4.5 kg to increase compression; the smallest $\tilde P$ is achieved by adjusting the length of the spring supports. 

\begin{figure}
\centerline{\includegraphics[width=0.9\linewidth]{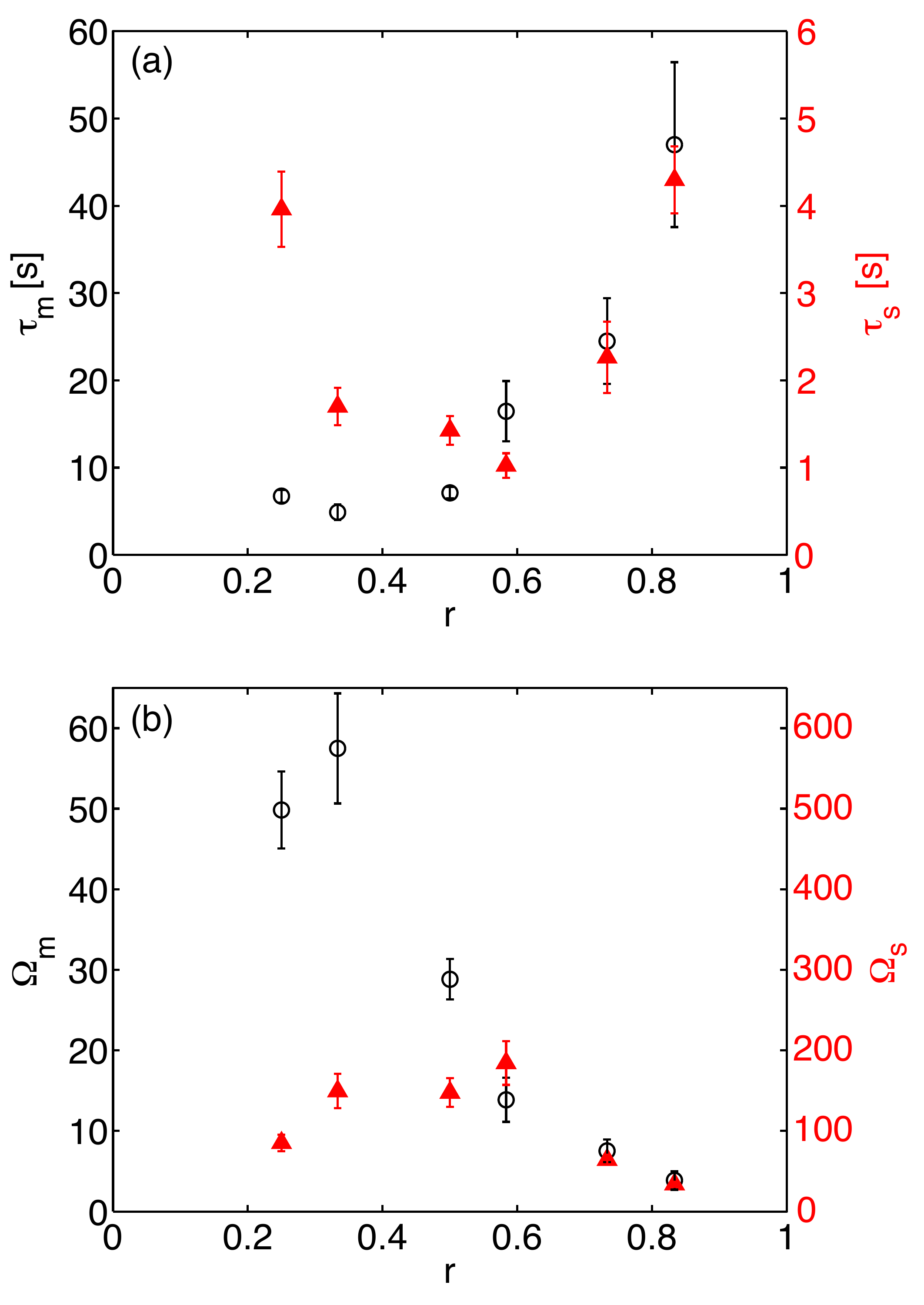}}
\caption{(a) Mixing timescale $\tau_m$ ($\circ$, left axis) and segregation timescale $\tau_s$ ($\blacktriangle$, right axis), as function of particle size ratio $r$. (b) Mixing rate $\Omega_m$ ($\circ$, left axis) and segregation rate $\Omega_s$ ($\blacktriangle$, right axis), as function of particle size ratio $r$. The error bars represent the standard error among at least 5 independent measurements. All data is collected at $\tilde P = 0.36$. \label{f:ratio} }
\end{figure}

We measure the mixing and segregation rates for six different particle size ratios with ${\tilde P} = 0.36$ (at least 5 runs each) and at six different pressures for $r=2/6$ and $r=5/6$ (at least 3 runs each). Fig.~\ref{f:ratio} depicts the mixing and segregation timescales and rates as a function of particle size ratio with pressure held constant. We observe that the mixing rate $\Omega_m$ decreases as particles become more similar in size ($r \rightarrow 1$). This corresponds to the expected kinetic-sieving behavior \citep{Williams-1976-SPM, Scott-1975-IPF, Savage-1988-PSS, Gray-2005-TPS} whereby small particles filter down through a fluctuating ``sieve'' of large particles. The smaller $r$ is, the more likely the small particles are to find voids to fall into. 

In contrast, we observe that the re-segregation process takes longer for both small and large $r$, as shown in Fig.~\ref{f:ratio}, whether measured as elapsed time or a rate  scaled by $\delta$. A maximum segregation rate is achieved near $r=3/6$: further reductions in the smaller particle size  slow the rate at which the system re-segregates.

\begin{figure}
\centerline{\includegraphics[width=0.9\linewidth]{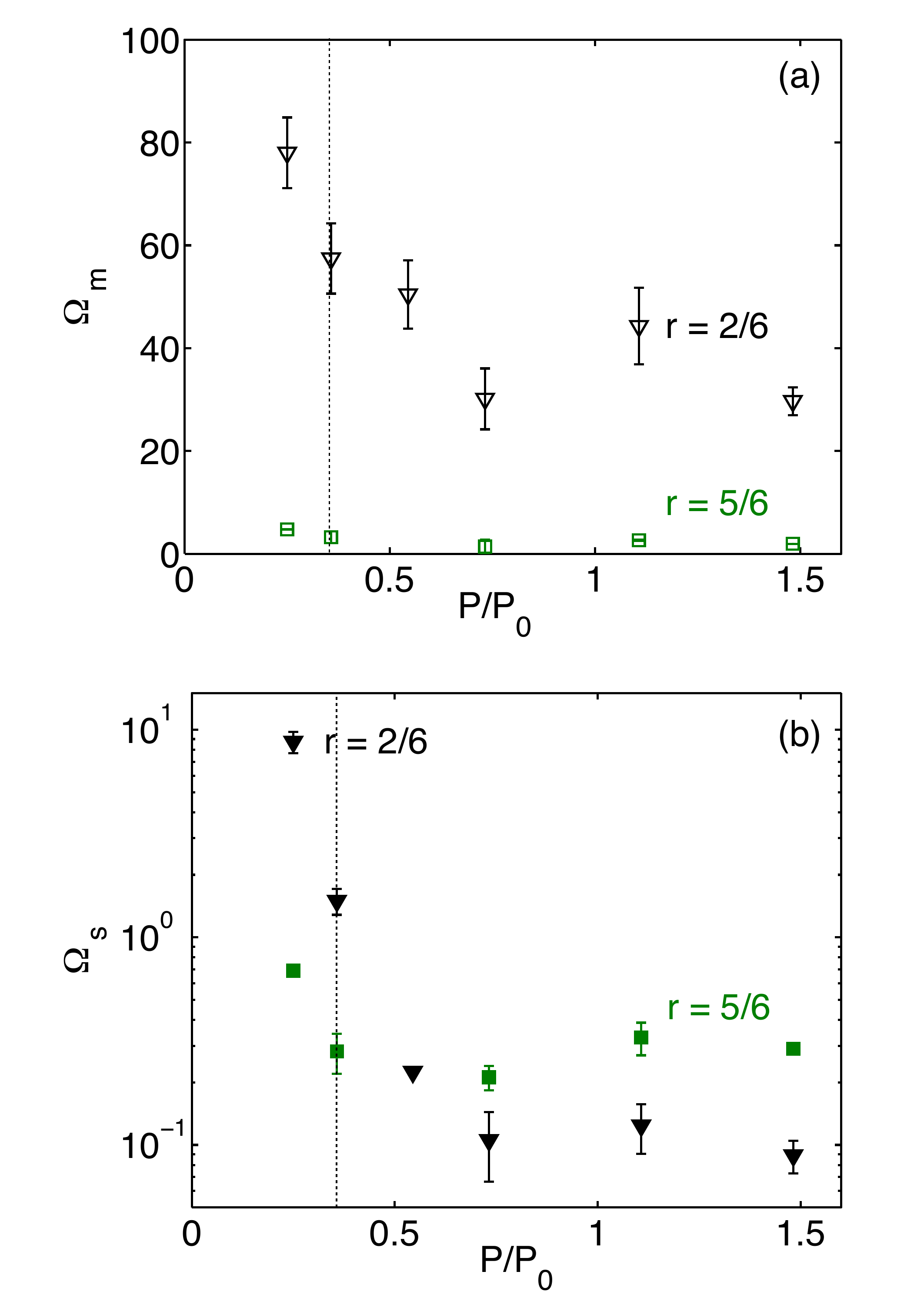}}
\caption{(a) Mixing rate $\Omega_m$ as a function of scaled pressure for $r=2/6$ ($\triangledown$) and $r=5/6$ ($\square$). (b) Segregation rate $\Omega_s$ as a function of scaled pressure for $r=2/6$ ($\blacktriangledown$) and $r=5/6$ ($\blacksquare$). The error bars represent the standard error among at least 3 independent measurements. Dashed line is $\tilde P = 0.36$, which coincides with data from Fig.~\ref{f:ratio}.}
\label{f:weight}
\end{figure}

In order to better understand this behavior, it is worthwhile to examine how $\Omega_s$ depends on the confining pressure on the system at both large and small $r$. As shown in Fig.~\ref{f:weight}, increasing $\tilde P$ decreases $\Omega_m$, and the pressure affects contrasting particle sizes (low $r$) more strongly than similar particle sizes. For $\Omega_s$ at low $r$, this effect is even more pronounced: a five-fold increase in pressure decreases the segregation rate by a factor of 100. This strong suppression of segregation with pressure causes an inversion in the $r$-dependence for ${\tilde P} \gtrsim 0.5$. Pressure has little effect on either rate as $r \rightarrow 1$.

These results provide three effects in need of explanation: (1) segregation rates display non-monotonic dependence on particle size ratio, (2) contrasting particle sizes are much more sensitive to pressure than similar particle sizes, and (3) mixing rates are much faster than segregation rates. The decrease in the segregation rate for small $r$ is particularly notable, since it is inconsistent with the predictions of kinetic sieving. The pressure-sensitivity of the system suggests looking at force chains \citep{Howell-1999-SF2,Mueth-1998-FDG} as an important factor for all three effects.

In simulations of granular materials in two dimensions, it is observed that force chains preferentially form through the larger particles as size ratio $r$ decreases \citep{Tsoungui-1998-PPS, Taboada-2005-RFT, Smart-2008-GMN, Voivret-2009-MFN}; this is likely related to the large particles' enhanced number of contacts. This unequal partitioning of force chains between large and small particles, were it to also be present for three-dimensional granular materials, could account for the first two effects. For small $r$, the presence of a large-particle-dominated force chain network at larger pressures could make it difficult for small particles to rearrange, thus slowing the segregation rate. As the particles become more similar in size (increasing $r$), such an imbalance would be smaller in magnitude. 

Another factor that could influence the anomalously low segregation rate of the system for small $r$ is the observed increase in packing fraction for mixtures of dissimilar particle sizes \citep{Kristiansen-2005-SRP}. If the experiments at low $r$ are denser, then they have less void space and this could slow their re-segregation.

We also observe a lack of reciprocity in the mixing and segregation mechanisms: a small particle falling though a mixture of mostly large particles (mixing) does not progress at the same rate as a large particle rising through a mixture of both large and small particles (segregation). Not only is there an approximately $10 \times$ difference in the associated rates (see Fig.~\ref{f:ratio}), but the segregation rates are much more pressure-dependent than the mixing rates. The mixing process can be more clearly associated with the void-filling mechanisms of kinetic sieving, which are apparently not strongly influenced by pressure. However, the segregation process requires large particles rising (called ``squeeze expulsion'' by \citep{Savage-1988-PSS}), which cannot be described by void filling. 

These experiments highlight the fact that granular segregation provides a sensitive probe of how  both the void space and the stress transmission influences the dynamics of the system. The pressure-dependency of the results suggest that volume-based descriptors of the state of granular systems \citep{Edwards-1989-TP} should be supplemented by information on the stresses \citep{Brujic-2003-JSS, Edwards-2005-FCE, Henkes-2007-ETS, Metzger-2008-HTC, Tighe-2008-EMF}. In the experiments described here, we are unable to measure either the void distributions or the force distributions for large-$P$ and small-$P$ cases, so we cannot disentangle the two effects. While local free volume distributions have recently been measured in three-dimensional systems \citep{Aste-2008-EGD}, little is yet known how such distributions are affected by pressure or shear. An improved understanding of the interplay between pressure and volume state variables will improve models of segregation.

\bigskip

\paragraph{Acknowledgments} The authors are grateful to David Fallest and Katherine Phillips for  initial hardware development and experiments, and to Michael Shearer, Lindsay May, Nico Gray and Kimberley Hill for fruitful discussions. This work has been supported by the National Science Foundation, under grants DMS-0604047 and DMR-0644743.



\end{document}